# 2016 Google Scholar Metrics released: a matter of languages... and something else

**Alberto Martín-Martín[1], Juan Manuel Ayllón[1], Enrique Orduña-Malea[2], Emilio Delgado López-Cózar[1]**

*[1] EC3 Research Group: Evaluación de la Ciencia y de la Comunicación Científica, Universidad de Granada (Spain)*
*[2] EC3 Research Group: Evaluación de la Ciencia y de la Comunicación Científica, Universidad Politécnica de Valencia (Spain)*

## ABSTRACT

The 2016 edition of Google Scholar Metrics was released on July 15th 2016. There haven't been any structural changes respect to previous versions, which means that most of its limitations still persist. The biggest changes are the addition of five new language rankings (Russian, Korean, Polish, Ukrainian, and Indonesian) and elimination of two other language rankings (Italian and Dutch). In addition, for reasons still unknown, this new edition doesn't include as many working paper and discussion paper series as previous editions.

Corresponding author
Emilio Delgado López-Cózar. edelgado@ugr.es

# 1. INTRODUCTION

We can only be delighted by the publication of the new edition of Google Scholar Metrics (GSM) (Thursday, July 14, 2016,6:44 PM) (Figure 1). These fifteen days of delay respect to the release of the previous version in 2015 (Thursday, June 25th, 2015, 12:16 PM) were starting to worry us, but we see now that these worries were unfounded. This year, GSM has been the last product for journal evaluation through citation analysis to be updated: the new editions of the Journal Citation Reports, Journal Metrics, and the SCImago Journal Rank were released in June.

**Figure 1. Top 19 publications in English according to Google Scholar Metrics 2016**

Top publications - English Learn more

| | Publication | h5-index | h5-median |
|---|---|---|---|
| 1. | Nature | 379 | 560 |
| 2. | The New England Journal of Medicine | 342 | 548 |
| 3. | Science | 312 | 464 |
| 4. | The Lancet | 259 | 418 |
| 5. | Cell | 224 | 339 |
| 6. | Chemical Society reviews | 224 | 329 |
| 7. | Journal of the American Chemical Society | 218 | 293 |
| 8. | Proceedings of the National Academy of Sciences | 215 | 286 |
| 9. | Advanced Materials | 201 | 301 |
| 10. | Angewandte Chemie International Edition | 198 | 276 |
| 11. | Journal of Clinical Oncology | 197 | 265 |
| 12. | Physical Review Letters | 196 | 282 |
| 13. | Chemical Reviews | 194 | 332 |
| 14. | Nano Letters | 192 | 270 |
| 15. | JAMA | 189 | 269 |
| 16. | Nucleic Acids Research | 184 | 345 |
| 17. | Energy & Environmental Science | 184 | 254 |
| 18. | ACS Nano | 180 | 243 |
| 19. | Nature Genetics | 179 | 267 |

As we said last year, we can only welcome that the American company has decided to keep supporting GSM, a free product which is also very different from traditional journal rankings. Competition is healthy, and scientists can only be pleased about this variety of search and ranking tools, especially when they are offered free of charge.

# 2. WHAT IS NEW IN GOOGLE SCHOLAR METRICS 2016?

There haven't been any structural changes in this new edition. The total number of publications that can be visualized in the 2016 rankings is 7,398. Now, however, since 1,664 of them (22.5%) are classified in more than one subject area, the number of unique publications is lower: 5,734. There are 12 language rankings, and inside the English rankings, there are 8 general categories and 262 unique subcategories. In this new edition, the subcategory “Corrosion”, which was available in the previous edition under the general category “Chemical & Material Sciences”, has been removed.

**Table 1. Number of disciplines by subject area. Google Scholar Metrics 2008-2012, 2009-2013, 2010-2014, 2011-2015**

| Subject Areas | Nº of disciplines | | | |
|---|---|---|---|---|
| | 2008-2012 | 2009-2013 | 2010-2014 | 2011-2015 |
| Physics & Mathematics | 26 | 24 | 24 | 24 |
| Chemical & Material Sciences | 20 | 19 | 19 | 18 |
| Engineering & Computer Science | 59 | 58 | 58 | 58 |
| Health & Medical Sciences | 72 | 69 | 69 | 69 |
| Life Sciences & Earth Sciences | 41 | 39 | 39 | 39 |
| Humanities, Literature & Arts | 28 | 26 | 26 | 26 |
| Business, Economics & Management | 17 | 16 | 16 | 16 |
| Social Sciences | 50 | 52 | 52 | 52 |

The main differences respect to last year's version is the inclusion of five additional language rankings (Russian, Korean, Polish, Ukrainian, and Indonesian) and the removal of two language rankings: Italian, and Dutch (Figure 2). The addition of new language rankings is welcome as they enrich the product. That's why we don't understand why they decided to remove the Italian and Dutch rankings.

**Figure 2. Screenshots of the left-side navigation bar in Google Scholar Metrics (2015vs2016)**

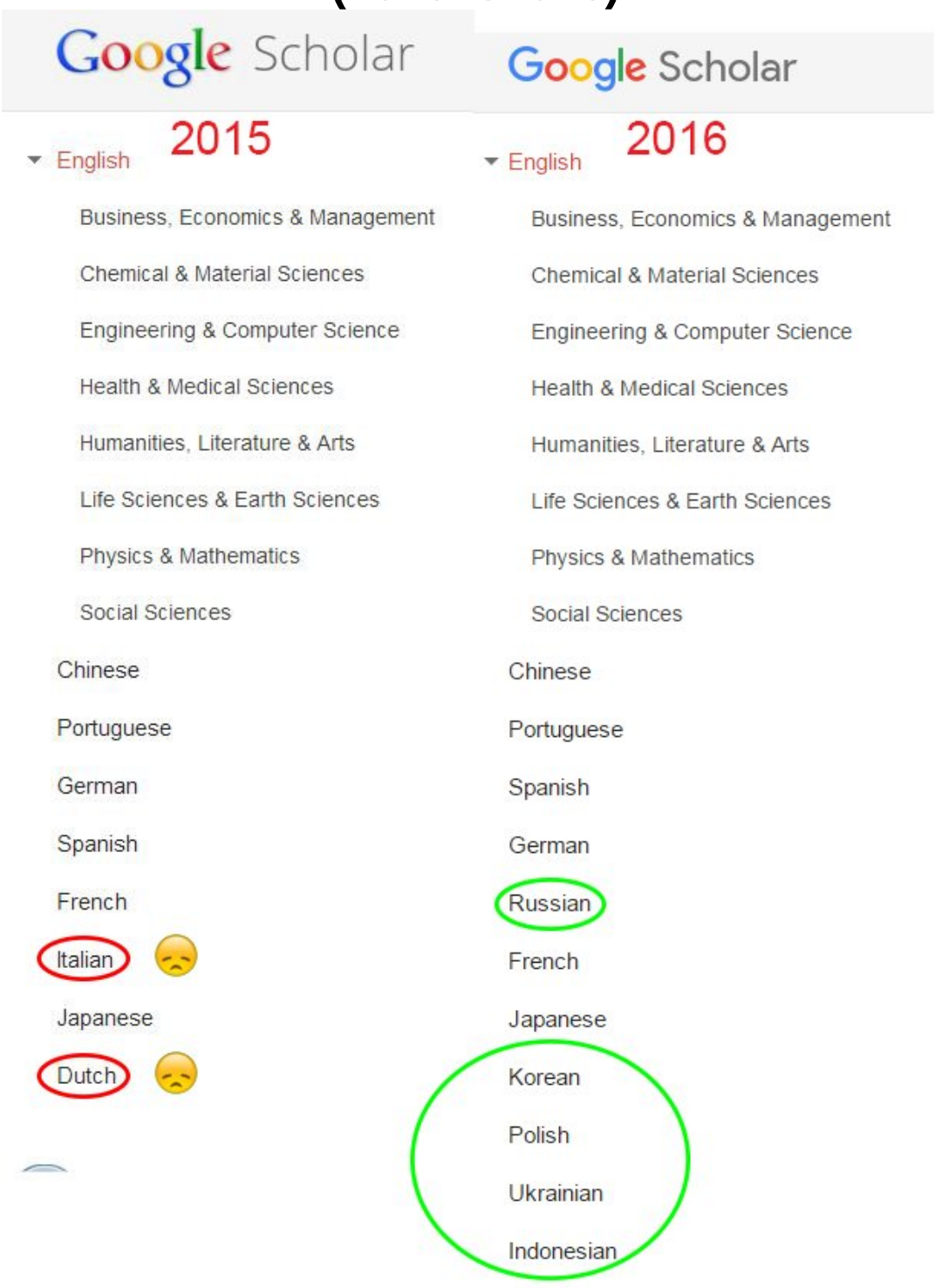

Another important change in this new version of Google Scholar Metrics is the removal of many Working Papers and Discussion Papers series. If users search "working papers", "discussion papers", "working paper", or "discussion paper" in GSM's search box, they will only get 7 results in total.

**Figure 3. Snapshots of various searches in Google Scholar Metrics 2016, trying to find as many working paper series as possible**

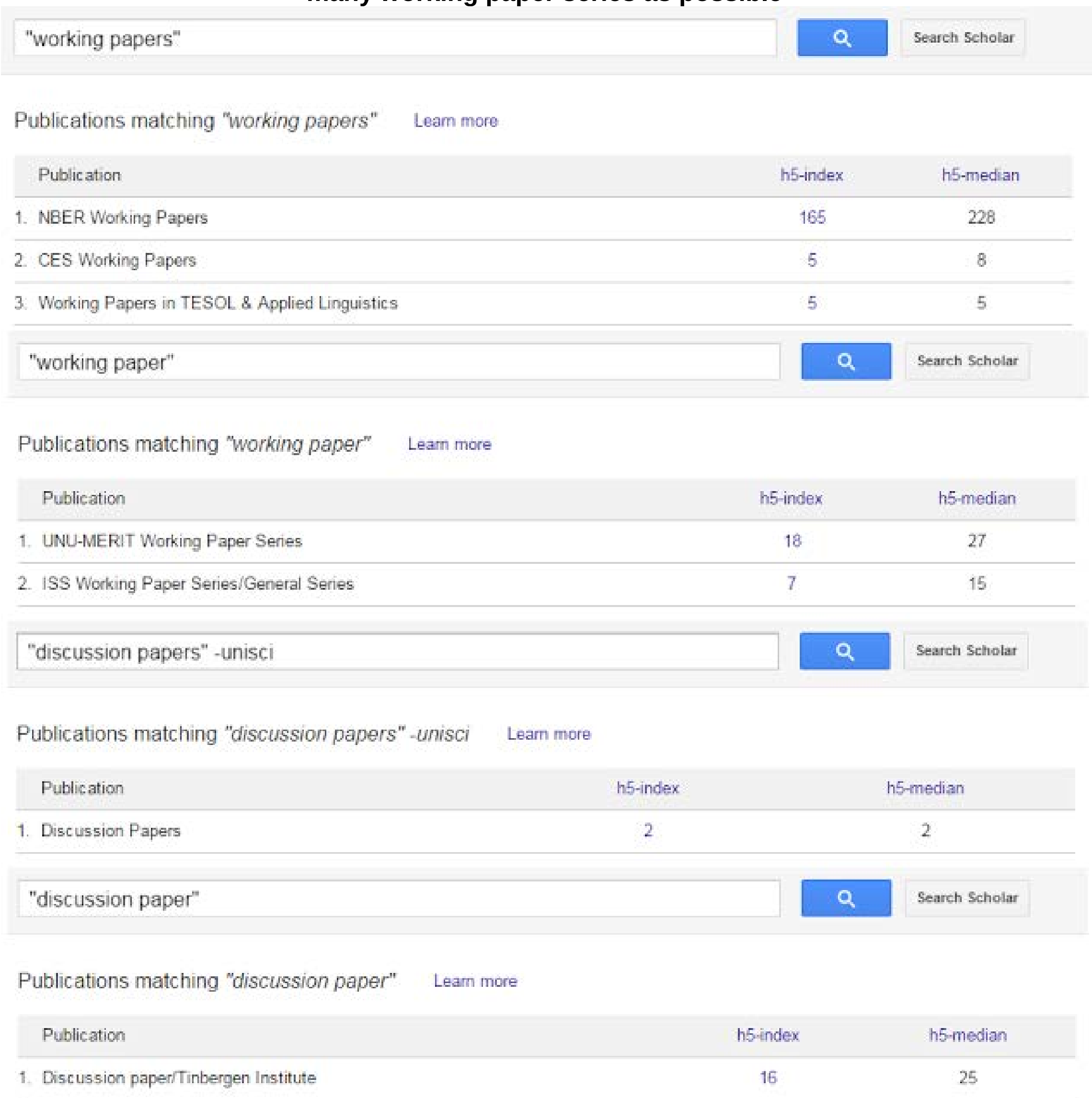

For example, in the previous edition of Scholar Metrics, the CEPR Discussion Papers (h5-index: 112) was ranked #4 in the general category Business, Economics & Management. This series even made it to the top 100 publications in the English ranking (93rd position). Similarly, the IZA Discussion Papers (h5-index: 82) was ranked #8 in the general category Business, Economics & Management. These two series are not to be found in the new edition of Scholar Metrics.

**Figure 4. CEPR and IZA Discussion Papers in Google Scholar Metrics 2015**

CEPR Discussion Papers

h5-index: **112** h5-median: **149**

**#1** Economic Policy
**#3** Economics
**#4** Business, Economics & Management

| Title / Author | Cited by | Year |
|---|---|---|
| Growth in a Time of Debt<br>CM Reinhart, K Rogoff<br>CEPR Discussion Papers | 1323 | 2010 |
| Measuring Systemic Risk<br>VV Acharya, LH Pedersen, T Philippon, MP Richardson<br>CEPR Discussion Papers | 856 | 2012 |

IZA Discussion Papers

h5-index: **82** h5-median: **124**

**#1** Human Resources & Organizations
**#8** Business, Economics & Management

| Title / Author | Cited by | Year |
|---|---|---|
| Estimating the Technology of Cognitive and Noncognitive Skill Formation<br>F Cunha, JJ Heckman, S Schennach<br>IZA Discussion Papers | 633 | 2010 |
| Personality Psychology and Economics<br>M Almlund, AL Duckworth, JJ Heckman, T Kautz<br>IZA Discussion Papers | 424 | 2011 |

One might think this has been caused by a change in GSM's inclusion policies [1]. They may have decided to remove all working papers and discussion papers, but if that were the case, they shouldn't have included other working paper series, like NBER Working Papers, currently #1 in the general category Business, Economics & Management, and also #1 in the subcategory Economics. They have also maintained all the subcategories available at arXiv. This is clearly inconsistent.

**Figure 5. Ranking of top publications in the general category "Business, Economics & Management" according to Google Scholar Metrics (2015 vs 2016)**

Top publications - Business, Economics & Management Learn more **2015**

| Publication | h5-index | h5-median |
|---|---|---|
| 1. NBER Working Papers | 163 | 233 |
| 2. The American Economic Review | 129 | 196 |
| 3. Journal of Financial Economics | 113 | 173 |
| 4. CEPR Discussion Papers | 112 | 149 |
| 5. The Journal of Finance | 108 | 172 |
| 6. Review of Financial Studies | 101 | 162 |
| 7. The Quarterly Journal of Economics | 93 | 170 |
| 8. IZA Discussion Papers | 82 | 124 |
| 9. Academy of Management Journal | 78 | 115 |
| 10. Management Information Systems Quarterly | 74 | 138 |
| 11. Journal of Management | 73 | 128 |
| 12. Review of Economics and Statistics | 73 | 123 |
| 13. Journal of Banking & Finance | 73 | 95 |
| 14. Econometrica | 72 | 130 |
| 15. Tourism Management | 72 | 99 |
| 16. Strategic Management Journal | 71 | 113 |
| 17. The Journal of Economic Perspectives | 71 | 110 |
| 18. Journal of Business Ethics | 70 | 88 |
| 19. The Review of Economic Studies | 69 | 117 |
| 20. International Journal of Production Economics | 69 | 95 |

Top publications - Business, Economics & Management Learn more **2016**

| Publication | h5-index | h5-median |
|---|---|---|
| 1. NBER Working Papers | 165 | 228 |
| 2. The American Economic Review | 137 | 218 |
| 3. Journal of Financial Economics | 121 | 171 |
| 4. The Journal of Finance | 108 | 176 |
| 5. Review of Financial Studies | 103 | 161 |
| 6. The Quarterly Journal of Economics | 93 | 167 |
| 7. Journal of Management | 79 | 120 |
| 8. Tourism Management | 79 | 109 |
| 9. Journal of Business Ethics | 78 | 104 |
| 10. Management Science | 78 | 100 |
| 11. Academy of Management Journal | 77 | 125 |
| 12. International Journal of Production Economics | 76 | 96 |
| 13. Management Information Systems Quarterly | 75 | 134 |
| 14. Econometrica | 74 | 128 |
| 15. The Journal of Economic Perspectives | 73 | 128 |
| 16. Review of Economics and Statistics | 72 | 119 |
| 17. World Development | 72 | 104 |
| 18. Journal of Business Research | 72 | 95 |
| 19. Strategic Management Journal | 71 | 104 |
| 20. Journal of Banking & Finance | 71 | 96 |

## 3. ERRORS

Apart from these differences, Google has just updated the data, which means that some of the limitations outlined in previous studies still persist [2-7]: the visualization of a limited number of publications (100 for those that are not published in English), the lack of categorization by subject areas and disciplines for non-English publications, and normalization problems (unification of journal titles, problems in the linking of documents, and problems in the search and retrieval of publication titles).

**Figure 6. Examples of duplicate entries for the same journals in Google Scholar Metrics**

Duplicate titles

| | | |
|---|---|---|
| 1. Endoxa | 3 | 4 |
| 2. Endoxa: Series Filosóficas | 3 | 4 |
| 1. Revista Icade. Revista de las Facultades de Derecho y Ciencias Económicas y Empresariales | 3 | 4 |
| 2. Icade: Revista de las Facultades de Derecho y Ciencias Económicas y Empresariales | 2 | 3 |
| 1. Enfermería Intensiva | 10 | 12 |
| 2. Enfermeria intensiva/Sociedad Espanola de Enfermeria Intensiva y Unidades Coronarias | 4 | 6 |
| 1. Insula: revista de letras y ciencias humanas | 4 | 4 |
| 2. INSULA-REVISTA DE LETRAS Y CIENCIAS HUMANAS | 2 | 2 |
| 1. EGA-Revista De Expresion Grafica Arquitectonica | 4 | 5 |
| 2. EGA. Revista de Expresión Gráfica Arquitectónica | 4 | 5 |
| 1. Estudios Geológicos | 9 | 18 |
| 2. ESTUDIOS GEOLOGICOS-MADRID | 7 | 11 |
| 1. Historia y política: Ideas, procesos y movimientos sociales | 3 | 6 |
| 2. HISTORIA Y POLITICA | 2 | 3 |
| 1. Revista espanola de enfermedades digestivas: organo oficial de la Sociedad Espanola de Patologia Digestiva | 20 | 25 |
| 2. Revista Española de Enfermedades Digestivas | 18 | 22 |

There are three different entries for the Brazilian Journal of Anesthesiology.

**Figure 7. Duplicate entries for the journal "Brazilian Journal of Anesthesiology" in Google Scholar Metrics**

Brazilian Journal of Anesthesiology | Search Scholar

Publications matching *Brazilian Journal of Anesthesiology* Learn more

| Publication | h5-index | h5-median |
|---|---|---|
| 1. Brazilian Journal of Anesthesiology | 16 | 21 |
| 2. Brazilian Journal of Anesthesiology (English Edition) | 10 | 13 |
| 3. Brazilian Journal of Anesthesiology (Edicion en Espanol) | 3 | 5 |

One of the main sources of errors in GSM are the journals published in several languages. Journals published in their original native language and, at the same time, in English, are quite common. GSM has decided to create separate entries for each of the languages in which a journal is published.

**Figure 8. Duplicate entries in Google Scholar Metrics for journals that are published in several languages**

| Publication | h5-index | h5-median |
|---|---|---|
| 1. Actas Urológicas Españolas (English Edition) | 17 | 29 |
| 2. Actas Urológicas Españolas | 13 | 16 |
| 1. Revista Portuguesa de Pneumologia (English Edition) | 18 | 22 |
| 2. Revista Portuguesa de Pneumologia | 14 | 21 |
| 1. Revista Española de Cirugía Ortopédica y Traumatología (English Edition) | 7 | 9 |
| 2. Revista Española de Cirugía Ortopédica y Traumatología | 7 | 8 |
| 3. Revista Rol de Enfermería | 4 | 6 |
| 4. Revista de enfermería (Barcelona, Spain) | 4 | 5 |
| 1. Revista Española de Medicina Nuclear e Imagen Molecular (English Edition) | 10 | 13 |
| 2. Revista Española de Medicina Nuclear e Imagen Molecular | 9 | 11 |

This decision is arguable, but at the very least, it should be applied consistently to all journals. The journal Revista Española de Cardiología, however, received a different treatment: the Spanish and English versions were merged.

**Figure 9. Most cited documents published in Revista Española de Cardiología, according to Google Scholar Metrics**

Google Scholar

Revista Española de Cardiología

h5-index: 37 h5-median: 49

| Title / Author | Cited by | Year |
|---|---|---|
| Third universal definition of myocardial infarction<br>K Thygesen, JS Alpert, AS Jaffe, ML Simoons, BR Chaitman, HD White<br>REVISTA ESPANOLA DE CARDIOLOGIA 66 (2), 2551-2567 | 3075 | 2013 |
| Guidelines on the management of valvular heart disease (version 2012) The Joint Task Force on the Management of Valvular Heart Disease of the European Society of Cardiology (ESC) and the European Association for Cardio-Thoracic Surgery (EACTS)<br>A Vahanian, O Alfieri, F Andreotti, MJ Antunes, G Baron-Esquivias, ...<br>REVISTA ESPANOLA DE CARDIOLOGIA 66 (2), E1-E42 | 1952 | 2013 |
| 2015 ESC Guidelines for the Management of Acute Coronary Syndromes in Patients Presenting Without Persistent ST-segment Elevation.<br>M Roffi, C Patrono, JP Collet, C Mueller, M Valgimigli, F Andreotti, JJ Bax, ...<br>Revista española de cardiología (English ed.) 68 (12), 11 | 280 | 2015 |
| 2014 ESC/ESA Guidelines on Non-cardiac Surgery: Cardiovascular Assessment and Management<br>SD Kristensen, J Knuuti, A Saraste, S Anker, HE Bøtker, S De Hert, I Ford, ...<br>Revista Española de Cardiología (English Edition) 67 (12 | 200 | 2014 |
| Factores de riesgo cardiovascular en España en la primera década del siglo xxi: análisis agrupado con datos individuales de 11 estudios de base poblacional, estudio DARIOS<br>M Grau, R Elosua, AC de León, MJ Guembe, JM Baena-Díez, TV Alonso, ...<br>Revista Española de Cardiología 64 (4), 295-304 | 152 | 2011 |
| Obesidad y corazón<br>F López-Jiménez, M Cortés-Bergoderi<br>Revista Española de Cardiología 64 (2), 140-149 | 96 | 2011 |
| Epidemiología del síndrome coronario agudo en España: estimación del número de casos y la tendencia de 2005 a 2049<br>IR Dégano, R Elosua, J Marrugat<br>Revista Española de Cardiología 66 (6), 472-481 | 83 | 2013 |

In the case of Revista Española de Enfermedades Digestivas and Revista Portuguesa de Neumología, they weren't able to successfully separate the two versions, since both versions present articles in the original languages (Spanish and Portuguese, respectively), and English.

**Figure 10. Examples of journals published in several languages where Google Scholar hasn't been able to separate documents by language**

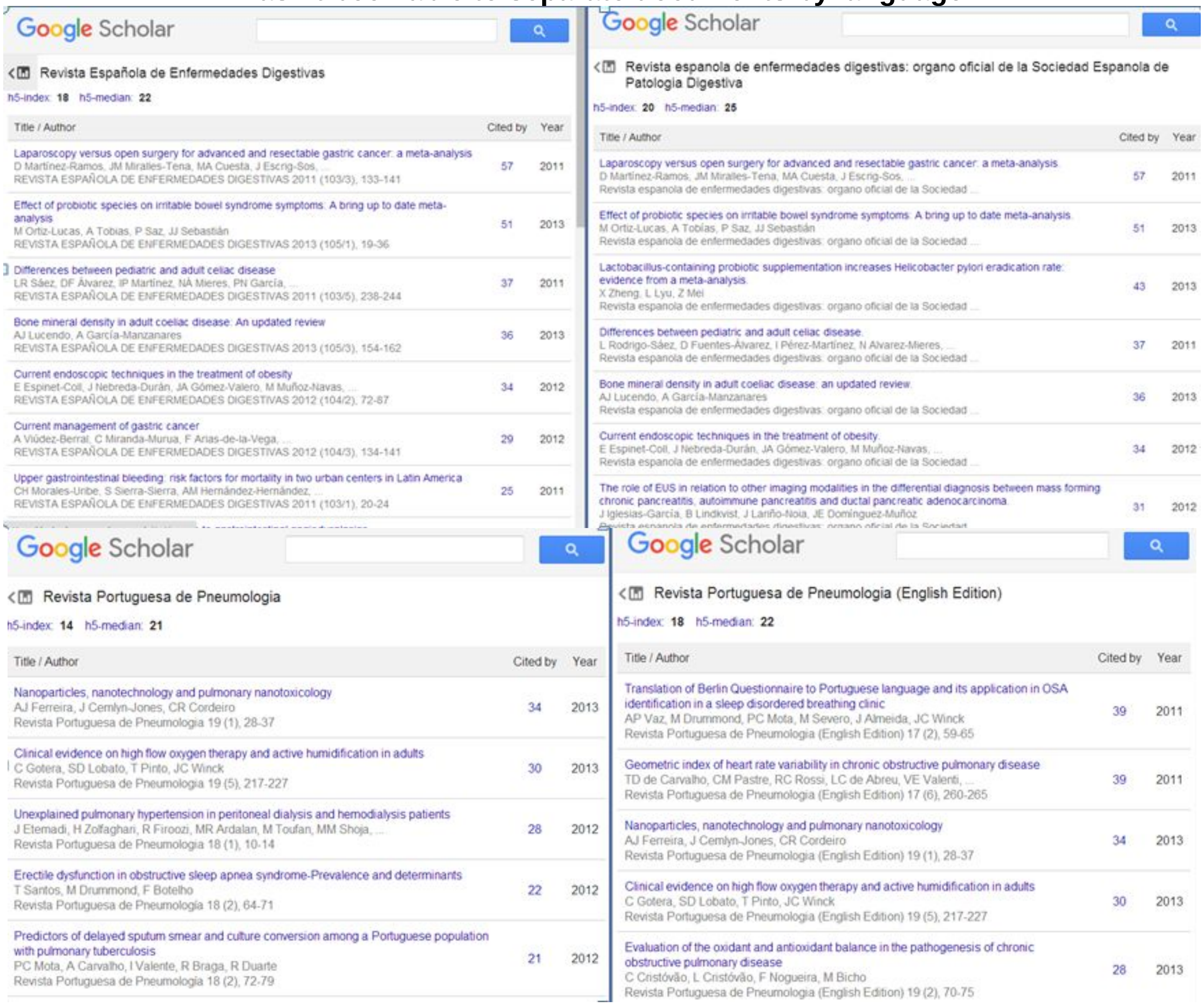

In the case of Giornale italiano di medicina del lavoro ed ergonomia, they only identified the English version, but not the Italian one.

**Figure 11. Searching "Giornale italiano di medicina del lavoro ed ergonomia" in Google Scholar Citations. Only the English version is found**

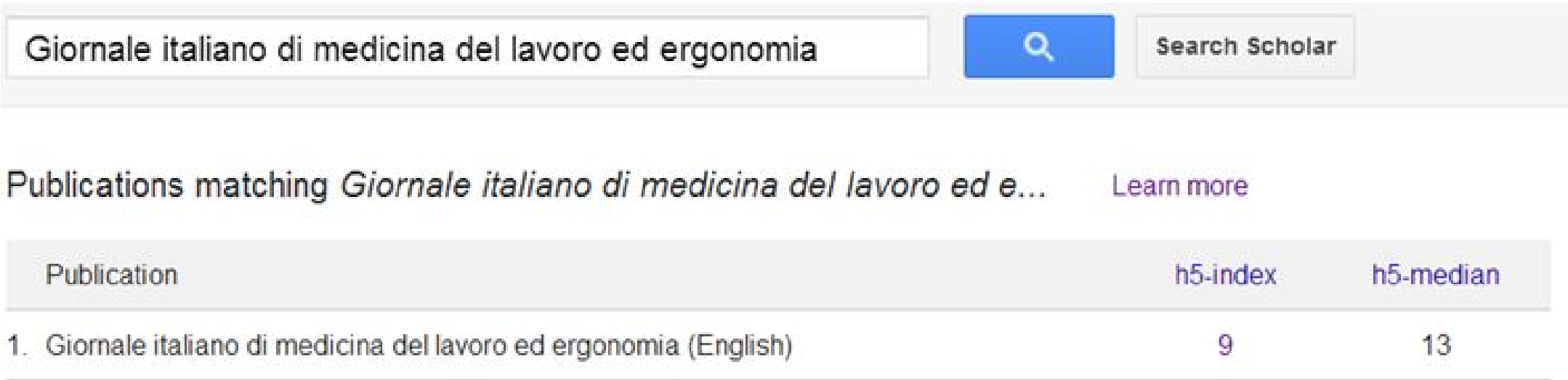

There are also several errores related to the correct linking of documents, which point to references or incorrect full-texts.

**Figure 12. Examples of documents with links that point to articles in other journals**

Enfermería Clínica

Índice h5: **11** Mediana h5: **15**

| Título / Autor | Citado por | Año |
|---|---|---|
| Concentric and eccentric: Muscle contraction or exercise?<br>J Padulo, J dal Pupo, G Laffaye, K Chamari<br>Enfermería Clínica 23 (4), 177-178 | 25 | 2013 |

Revista de Estudios Políticos

Índice h5: **7** Mediana h5: **11**

| Título / Autor | Citado por | Año |
|---|---|---|
| Presentación<br>AL Castillo, CC Montero<br>Revista de estudios políticos, 13-16 | 53 | 2014 |

Atencion primaria

h5-index: **19** h5-median: **26**

| Title / Author | Cited by | Year |
|---|---|---|
| Guía Española de la EPOC (GesEPOC). Tratamiento farmacológico de la EPOC estable<br>M Miravitlles, JJ Soler-Cataluña, M Calle, J Molina, P Almagro, ...<br>Atención Primaria 44 (7), 425-437 | 59 | 2012 |

In some cases, like the journal Nutrición Hospitalaria, we find dead links, links to the PDFs in Scielo, links to Dialnet, and links to the various repositories where authors have archived their articles. Probably for this reason the title of the journal presents up to three variants.

**Figure 13. Example of journal for which information is extracted from various sources (Scielo, Dialnet, repositories...)**

Nutrición hospitalaria

h5-index: **33** h5-median: **42**

| Title / Author | Cited by | Year |
|---|---|---|
| Prevalence and costs of malnutrition in hospitalized patients; the PREDyCES Study.<br>J Álvarez-Hernández, MP Vila, M León-Sanz, AG de Lorenzo, ...<br>Nutricion Hospitalaria 27 (4) | 111 | 2012 |
| Importance of a balanced omega 6/omega 3 ratio for the maintenance of health. Nutritional recommendations<br>IDELE DEL ÍNDICE, EDES DE UN BUEN, R NUTRICIONALES<br>Nutr Hosp 26 (2), 323-329 | 107 | 2011 |
| Prevalencia de sobrepeso y obesidad en adultos españoles<br>E Rodríguez-Rodríguez, B López-Plaza, AM López-Sobaler, RM Ortega<br>Nutrición Hospitalaria 26 (2), 355-363 | 85 | 2011 |
| Dietary fibre and cardiovascular health<br>FJ Sánchez-Muniz<br>Nutr Hosp 27 (1), 31-45 | 76 | 2012 |
| Energy expenditure: components and evaluation methods<br>ACP Volp, FCE De Oliveira, RD Moreira-Alves, EA Esteves, J Bressan<br>Nutrición hospitalaria: Organo oficial de la Sociedad española de nutrición ... | 73 | 2011 |

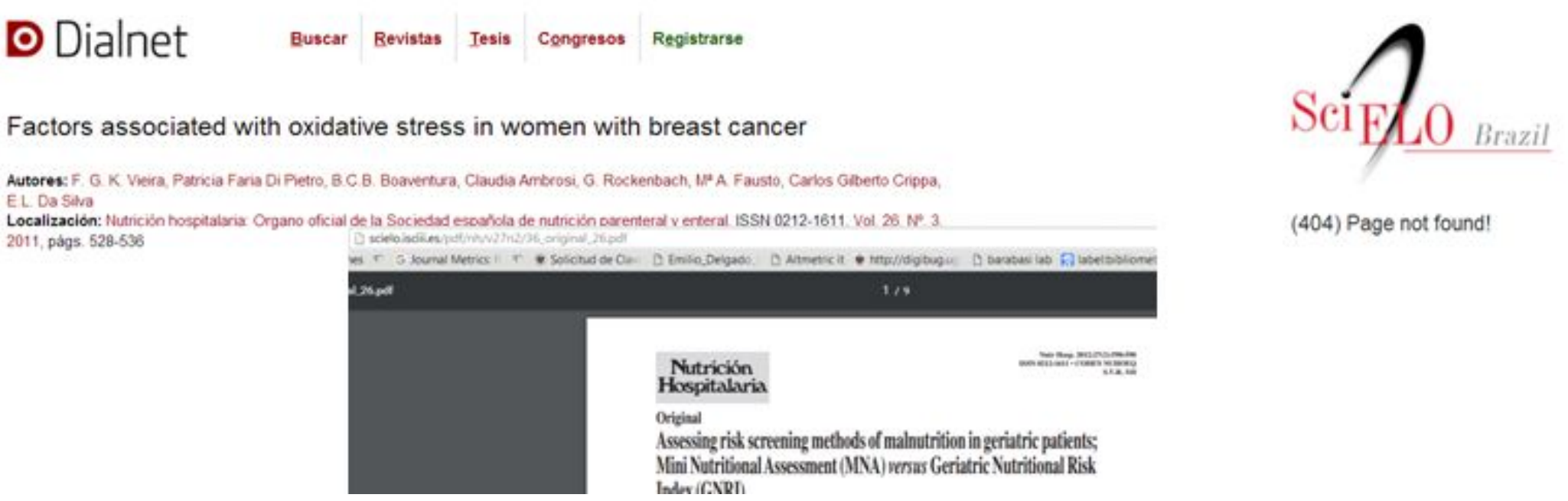

Over the years we have detected cases of journals that don't seem to meet all the criteria set by GSM to be included in this product (mainly the minimum of 100 articles published in the las five years), and nevertheless they are included. An example of this phenomenon is the journal Area Abierta, for which there are only 43 articles published between 2011 and 2015 indexed in Google Scholar, but still is included. Additionally, the most cited article in this journal is incorrect because it actually points to an article published in another journal.

**Figure 14. Example of journal which has published less than 100 articles in the previous 5-year period, and nevertheless is covered in Google Scholar Metrics**

Area Abierta

h5-index: **6** h5-median: **13**

| Title / Author | Cited by | Year |
|---|---|---|
| Presentación<br>FAZ Hernández<br>Area abierta 15 (1), 1-2 | 53 | 2015 |
| Los programas deportivos de la radio española en la red social Facebook: espacio de promoción, lugar de encuentro...¿ medidor de audiencia?/Sports programs of the Spanish radio in the social network Facebook: space of promotion, place of meeting...¿ audience measurement?<br>FJH Gutiérrez<br>Area Abierta, 1 | 20 | 2011 |
| Narrativa crossmedia en el discurso televisivo de Ciencia Ficción. Estudio de Battlestar Galactica (2003-2010)<br>MH PÉREZ, MDMG PÉREZ<br>Área Abierta, 4-4 | 19 | 2011 |
| Siglo XXI y monarquía. Propuestas para dinamizar la caracterización informativa del rey Juan Carlos I/21st Century and the Monarchy. Proposals to Dynamize the Informative Characterization of King Juan Carlos I<br>DB Ibáñez<br>Area Abierta 34 (3), 1 | 8 | 2013 |
| DE LA PUBLICIDAD ESPECTÁCULO1 A LOS VALORES EMOCIONALES: EL SECTOR DE LA ENERGÍA EN ESPAÑA/FROM THE" EVENT ADVERTISING" TO THE EMOTIONAL VALUES: THE SECTOR OF THE ENERGY IN SPAIN<br>AÁ Ruiz, MIR Moreno<br>Area Abierta, 1 | 6 | 2011 |
| Entre la huella y el índice: relecturas contemporáneas de André Bazín/From Between the trace and the index: Contemporary Readings of André Bazin<br>LE Verano, EC Álvarez<br>Area Abierta, 1 | 6 | 2012 |

The journal Investigaciones de Historia Económica presents a similar case: this journal doesn't publish the minimum 100 original articles in the last five years, and still it is included in GSM. If we search articles published by this journal in Google Scholar, we see that this journal publishes a high amount of book reviews. Probably, these reviews were considered as articles when the data was computed.

**Figure 15. Example of journal that has probably been included in GSM because it publishes many book reviews. Without them, it doesn't reach the minimum 100 publications in the previous 5-year period**

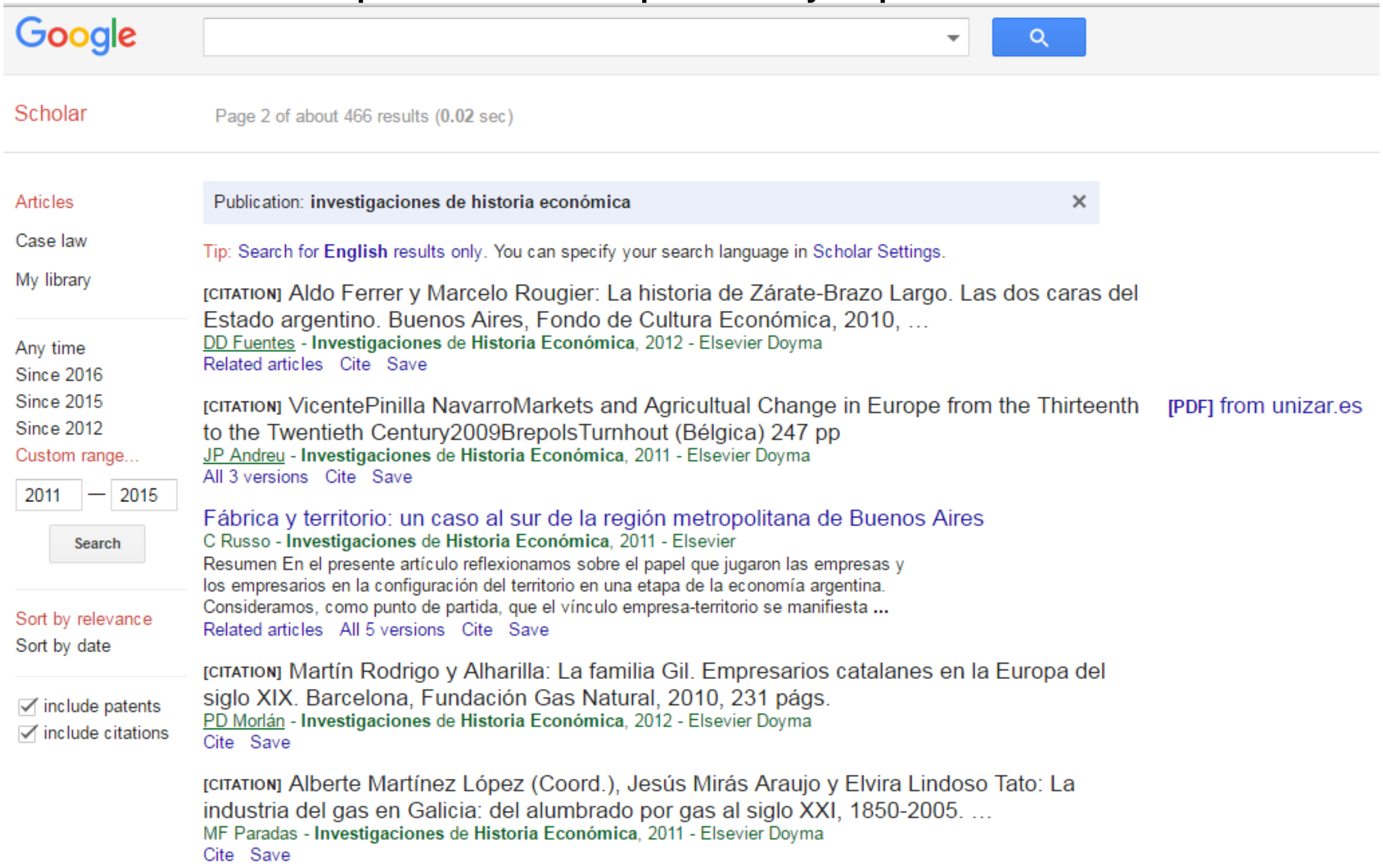

Lastly, it should be reminded that journals not always present a uniform typographic design in their titles or the titles of the articles.

**Figure 16. Example of articles with different font case settings: some are all in uppercase, and some use conventional rules**

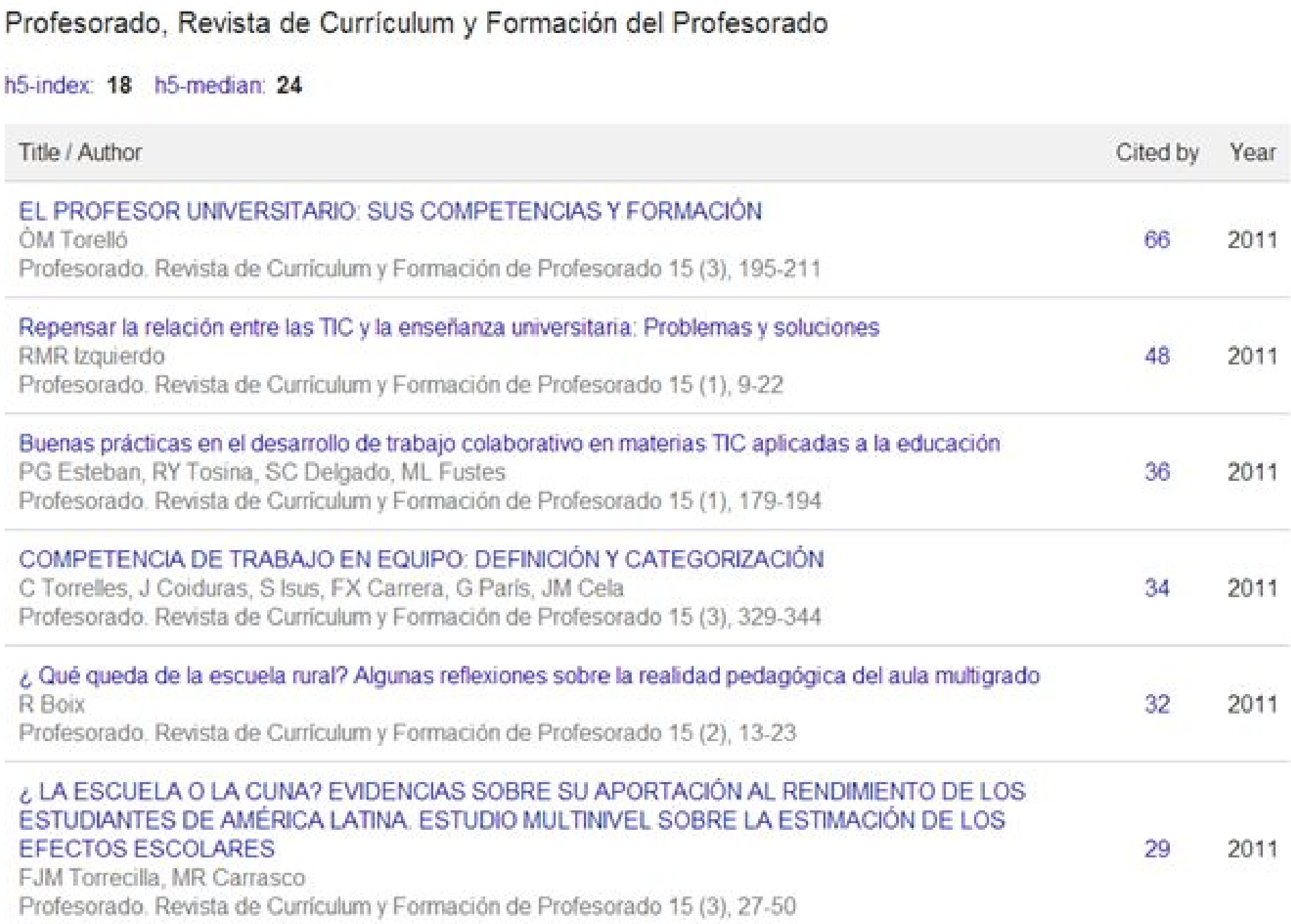
Profesorado, Revista de Currículum y Formación del Profesorado

h5-index: **18** h5-median: **24**

| Title / Author | Cited by | Year |
|---|---|---|
| EL PROFESOR UNIVERSITARIO: SUS COMPETENCIAS Y FORMACIÓN<br>ÒM Torelló<br>Profesorado. Revista de Currículum y Formación de Profesorado 15 (3), 195-211 | 66 | 2011 |
| Repensar la relación entre las TIC y la enseñanza universitaria: Problemas y soluciones<br>RMR Izquierdo<br>Profesorado. Revista de Currículum y Formación de Profesorado 15 (1), 9-22 | 48 | 2011 |
| Buenas prácticas en el desarrollo de trabajo colaborativo en materias TIC aplicadas a la educación<br>PG Esteban, RY Tosina, SC Delgado, ML Fustes<br>Profesorado. Revista de Currículum y Formación de Profesorado 15 (1), 179-194 | 36 | 2011 |
| COMPETENCIA DE TRABAJO EN EQUIPO: DEFINICIÓN Y CATEGORIZACIÓN<br>C Torrelles, J Coiduras, S Isus, FX Carrera, G París, JM Cela<br>Profesorado. Revista de Currículum y Formación de Profesorado 15 (3), 329-344 | 34 | 2011 |
| ¿ Qué queda de la escuela rural? Algunas reflexiones sobre la realidad pedagógica del aula multigrado<br>R Boix<br>Profesorado. Revista de Currículum y Formación de Profesorado 15 (2), 13-23 | 32 | 2011 |
| ¿ LA ESCUELA O LA CUNA? EVIDENCIAS SOBRE SU APORTACIÓN AL RENDIMIENTO DE LOS ESTUDIANTES DE AMÉRICA LATINA. ESTUDIO MULTINIVEL SOBRE LA ESTIMACIÓN DE LOS EFECTOS ESCOLARES<br>FJM Torrecilla, MR Carrasco<br>Profesorado. Revista de Currículum y Formación de Profesorado 15 (3), 27-50 | 29 | 2011 |

Having said that, there are fewer errors than in previous years.

## 4. SUGGESTIONS

In our previous studies, we have described again and again the underlying philosophy embedded in all of Google's academic products. These products have been created in the image and likeness of Google's general search engine: fast, simple, easy to use, understand and calculate?, and last but not least, accessible to everyone free of charge. GSM follows all these precepts, and it is, in the end, nothing more than:

- A hybrid between a bibliometric tool (indicators based on citation counts), and a bibliography (a list of highly cited documents, and of the documents that cite them).
- It offers a simple, straightforward journal classification scheme (although it also includes some conferences and repositories).
- It is based on two basic bibliometric indicators (the h index, and the median number of citations for the articles that make up the h index).
- It covers a single five-year time frame (the current one being 2011-2015).
- It uses rudimentary journal inclusion criteria, namely: publishing at least 100 articles during the last five-year period, and having received at least one citation.
- It provides lists of publications according to the language their documents are written in. For all of them, except for English publications (these are a total of 11: Chinese, Portuguese, German, Spanish, French, Japanese, Russian, Korean, Polish, Ukrainian and Indonesian) it offers lists of only 100 titles: those with the higher h index. For English publications, however, it shows a total of 4737 different publications, grouped in 8 subject areas. For each publication, it shows the titles of the documents whose citations contribute to the h index, and for each one of these documents, in turn, the titles of the documents that cite them.
- It provides a search feature that, for any given set of keywords, will retrieve a list of 20 publications whose titles contain the selected keywords. In the cases where there are more than 20 publications that satisfy the query, only the first 20 results, those with a higher h index, will be displayed.
- It doesn't perform any kind of quality control in the indexing process nor in the information visualization process.

To sum up, GSM is a minimalist information product with few features, closed (it cannot be customized by the user), and simple (navigating it only takes a few clicks). If GSM wants to improve as a bibliometric toolit should incorporate a wider range of features. At the very least, it should:

- Display the total number of publications indexed in GSM, as well as their countries and language of publication. Our estimations lead us to believe that this figure is probably higher than 40,000 [8]. In the case of Spain, there are over 1,000 publications indexed, which make up about 45% of the total number of academic publications in Spain [9-11].

- Provide some other basic and descriptive bibliometric indicators, like the total number of documents published in the publications indexed in GSM, and the total number of citations received in the analysed time frame. These are the two essential parameters that make it possible to assess the reliability and accuracy of any bibliometric indicator. Other indicators could be added in order to elucidate other issues like self-citation rates, impact over time (immediacy index), or to normalize results (citation average).
- Provide the complete list of documents of any given publication that have received n citations and especially those that have received 0 citations. This would allow us to verify the accuracy of the information provided by this product. It is true, much to Google's credit, that this information could be extracted, though not easily?, from Google Scholar.
- Provide a detailed list of the conferences and repositories included in the product. The statement Google makes about including some conferences in the Engineering & Computer Science area, and some document collections like the mega-repositories arXiv, RePec and SSRN, is much too vague.
- Define the criteria that has been followed for the creation of the classification scheme (areas and disciplines), and the rules and procedures followed when assigning publications to these areas and disciplines.
- Enable the selection of different time frames for the calculation of indicators and the visualization and sorting of publications. The significant disparities in publishing processes and citation habits between areas (publishing speed, pace of obsolescence) require the possibility to customize the time frame according to the particularities of any given subject area.
- Enable access to previous versions of Google Scholar Metrics (2007-2011, 2008-2012, 2009-2013, 2010-2014) to ensure that it is possible to assess the evolution of publications over time. Moreover, they could dare venture into the unknown and do something no one else has done before: a dynamic product, with indicators and rankings updated in real-time, just as Google Scholar does.
- Enable browsing publications by language, country and discipline, and directly display all results for these selections.
- Remove visualization restrictions: currently 100 results for each language and 20 for each discipline or keyword search.
- Enable the visualization of results by country of publication and by publisher.
- Enable sorting results according to various criteria (publication title, country, language, publishers), as well as according to other indicators (h index, h median, number of documents per publication, number of citations, self-citation rate…).
- Enable searching not only by publication title, but also by country and language of publication.
- Enable an option for exporting global results, as well as results by discipline, or those of a custom query.

- Enable an option for reporting errors detected by users, so they can be fixed (duplicate titles, erroneous titles, incorrect links, deficient calculations…).
- Lastly, reducing the minimum number of articles published in the last 5 years from 100 to 50 might be a good idea. 20 articles per year is not a difficult goal for journals written in English, especially in areas like natural sciences and health. However, there are many local journals published in non-English-speaking countries, especially in the Arts & Humanities, that just can't reach that amount of articles.

"Dixit two years ago"